\documentclass[prl,twocolumn,aps,a4paper,superscriptaddress]{revtex4}
\usepackage{dcolumn}
\usepackage{amsmath}
\usepackage{graphicx}
\usepackage{latexsym}
\usepackage{amsfonts}
\usepackage{amssymb}
\usepackage{dsfont}

\usepackage{amssymb,amsmath,stmaryrd}
\usepackage{times}
\usepackage{float}
\usepackage{pst-node}
\usepackage[nouppercase]{scrpage2}
\usepackage{makeidx}
\usepackage{bbm}
\usepackage{cancel}
\usepackage{amsthm}
\usepackage{textgreek}

\DeclareGraphicsExtensions{.pdf,.gif,.jpg}

\newcommand{\be}{\begin{equation}}
\newcommand{\ee}{\end{equation}}
\newcommand{\bea}{\begin{eqnarray}}
\newcommand{\eea}{\end{eqnarray}}

\def\a{\alpha}

\def\g{\gamma}
\def\G{\Gamma}
\def\d{\delta}
\def\D{\Delta}
\def\e{\epsilon}
\def\ve{\varepsilon}

\def\th{\theta}

\def\l{\lambda}

\def\m{\mu}

\def\c{\xi}

\def\p{\pi}

\def\s{\sigma}

\def\vf{\varphi}

\def\w{\omega}
\def\W{\Omega}
\def\q{\psi}

\def\callN{\mbox{$\mathcal{N}$}}

\def\callT{\mbox{$\mathcal{T}$}}

\def\iif{\infty}

\def\1op{\hat{\mathbbm{1}}}

\begin{document}

\title{Floquet topological phase of nondriven $p$-wave nonequilibrium excitonic insulators}

\author{E. Perfetto}
\affiliation{Dipartimento di Fisica, Universit\`{a} di Roma Tor Vergata,
Via della Ricerca Scientifica 1, 00133 Rome, Italy}
\affiliation{INFN, Sezione di Roma Tor Vergata, Via della Ricerca
  Scientifica 1, 00133 Rome, Italy}
\author{G. Stefanucci}
\affiliation{Dipartimento di Fisica, Universit\`{a} di Roma Tor Vergata,
Via della Ricerca Scientifica 1, 00133 Rome, Italy}
\affiliation{INFN, Sezione di Roma Tor Vergata, Via della Ricerca
  Scientifica 1, 00133 Rome, Italy}

\begin{abstract}
The nontrivial topology of {\it p}-wave superfluids make these 
systems attractive candidate in information technology.
In this work we report on the topological state of a {\it p}-wave 
nonequilibrium excitonic insulator (NEQ-EI) and show how to steer a 
nontopological band-insulator with bright $p$-excitons toward this state by a suitable laser 
pulse, thus achieving a dynamical topological phase transition. 
The underlying mechanism behind the transition is the broken 
gauge-symmetry of the NEQ-EI which causes self-sustained
persistent oscillations of the excitonic condensate and hence 
a Floquet topological state for high enough exciton densities.
We show the formation of Floquet Majorana modes at the boundaries of 
the open system and discuss topological spectral signatures for ARPES 
experiments.
We emphasize that the topological properties of a {\it p}-wave NEQ-EI 
arise exclusively from the electron-hole Coulomb interaction as the 
system is not driven by external fields.
%

\end{abstract}

\maketitle

A quantum system with nontrivial bulk topological properties admits 
localized single-particle states at the system 
edges~\cite{Read_PhysRevB.61.10267,Hasan_RevModPhys.82.3045,Qi_RevModPhys.83.1057}. 
Existence of well-defined {\em quasi-particles}  is of course a 
prerequisite for the bulk--edge correspondence to have physical 
relevance. In fact, most topological invariants are constructed from a 
quasi-particle hamiltonian which is treated either as 
``non-interacting'', i.e., independent of the charge distribution, or 
at a ``mean-field'' level. Mean-field hamiltonians introduce an 
appealing twist 
in the topological characterization since, at fixed 
external potentials, they 
depend on the (self-consistent) charge distribution of the 
stationary state. Thus, in principle, a quantum system can change its 
topological properties upon a transition from an excited state to 
another. Furthermore, the possibility of self-sustained, i.e., 
not driven by external fields, 
oscillatory solutions extends the class of topological invariants to the 
Floquet realm~\cite{Rechtsman2013,Cayssol_2013,Kitagawa_PhysRevB.82.235114,Rudner_PhysRevX.3.031005}. 

Non-equilibrium (NEQ) excitonic insulators (EI) are excited states of 
band-insulator (BI) mean-field hamiltonians giving rise to a 
self-sustained
oscillating order parameter, i.e., the excitonic condensate 
(EC)~\cite{Ostreich_1993,TriolaPRB2017,PertsovaPRB2018,Hanai2016,HanaiPRB2017,Hanai2018,Becker_PhysRevB.99.035304,Yamaguchi_NJP2012,Yamaguchi_PhysRevLett.111.026404,Hannewald-Bechstedt_2000,PSMS.2019}. 
In this Letter we show that a {\em p}-wave NEQ-EI undergoes a 
topological transition with increasing the EC density, leading to the 
formation  of Floquet Majorana edge modes~\cite{Jiang_PhysRevLett.106.220402}.
We further demonstrate that the Floquet topological {\em p}-wave 
NEQ-EI can be built up 
in real time by laser pulses of finite duration provided that 
$p$-excitons exist and are optically active. As 
the initial BI ground-state has vanishing EC density and trivial 
topology the system experiences a {\it dynamical} phase transition 
(from BI to non-topological NEQ-EI) followed by a topological one. The density of topological defects predicted by the
Kibble-Zurek mechanism~\cite{Kibble_1976,Zurek1985} for external 
drivings of finite 
duration~\cite{Sengupta_PhysRevLett.100.077204,Mondal_PhysRevB.78.045101,Bermudez_PhysRevLett.102.135702,Bermudez_2010,Lee_PhysRevB.92.035117,Defenu_PhysRevB.100.184306} 
can indeed be made sufficiently 
small to preserve the topological character of the final state.
Unique spectral signatures for experimental ARPES investigations are also discussed. In particular, 
at the topological transition point the spectrum 
becomes gapless and the {\em p}-wave NEQ-EI turns into a Dirac semimetal.





Nonvanishing topological 
invariants~\cite{Read_PhysRevB.61.10267,Qi_RevModPhys.83.1057} and 
existence of Majorana edge modes~\cite{Kitaev_2001}
in quantum matter with a {\it p-}wave 
symmetry--broken ground state have been recently reported for 
superconductors~\cite{Volovik1999,Ivanov_PhysRevLett.86.268,Kallin_2016,Sato_2017},
superfluids of ultracold atomic 
gases~\cite{GURARIE20072}, insulators~\cite{Hasan_RevModPhys.82.3045} and 
excitonic insulators~\cite{Pikulin_PhysRevLett.112.176403,Wang2019}. 
In nonequilibrium and nondriven conditions, however, a nontrivial 
topology  has so far been found only in the mean-field Floquet 
hamiltonian of a {\it p-}wave superfluid~\cite{Foster_PRL2014}. 


The simplest description of a NEQ-EI is provided by a spinless 
one-dimensional hamiltonian 
with a single valence and conduction bands separated by a direct gap 
of magnitude $\e_{g}$~\cite{PSMS.2019}:
\be
\hat{H}=\sum_{\a j}(-)^{\a}[V\hat{\q}^{\dag}_{\a 
j}\hat{\q}_{\a j+1}-(2V+\e_{g}/2)\hat{n}_{\a j}] + 
\sum_{ij}U_{ij}\hat{n}_{vi}\hat{n}_{cj},
\label{minmodham}
\ee
where $\hat{\q}_{v j}$ ($\hat{\q}_{c j}$) annihilates a valence 
(conduction) electron in the $j$-th cell, $\hat{n}_{\a j}\equiv\hat{\q}^{\dag}_{\a 
j}\hat{\q}_{\a j}$ and $(-)^{\a}=1,-1$ for $\a=v,c$ (the hopping 
integral $V$ is chosen positive). The hamiltonian is invariant under the 
``local'' gauge symmetry $\hat{\q}_{\a j}\to 
e^{i\th_{\a}}\hat{\q}_{\a j}$ 
associated to the commutation relation $[\hat{H},\hat{N}_{\a}]=0$, 
with $\hat{N}_{\a}=\sum_{j}\hat{n}_{\a j}$.
For large enough $U_{ij}$ the ground state is a BI with a filled 
(empty) valence (conduction) band. Charge neutral excited states with $N_{c}=1$ 
can be calculated by solving the Bethe-Salpeter equation (BSE).
For short-range interactions, e.g., $U_{ij}=\d_{ij}U$, the BSE admits 
only one discrete solution corresponding to an
{\it s}-wave (even) exciton~\cite{PSMS.2019}. 
A Rydberg-like series, and hence {\it p}-wave (odd) excitons, appears 
with long-range interactions such as the soft-Coulomb one: 
$U_{ij}=U/\sqrt{|i-j|^{2}+1}$. Henceforth we express all energies  in 
units of $\e_{g}$ and choose $U=2V=1$. Then, the BSE admits
multiple excitonic solutions, the two lowest having energy
$\e_{\mathrm{x}}^{s}=0.40$ ({\it s}-wave) and 
$\e_{\mathrm{x}}^{p}=0.82$ ({\it p}-wave) above the valence band 
maximum. Charge-neutral excited states with a finite density in 
the conduction band will be treated in the mean-field approximation.

The lowest-energy excited state of $\hat{H}$ with a finite density 
of conduction electrons and valence holes equals the ground state of 
the NEQ gran-canonical Hamiltonian 
$\hat{H}_{\rm NEQ-GC}\equiv 
\hat{H}-\m_{v}\hat{N}_{v}-\m_{c}\hat{N}_{c}$, where $\m_{\a}$ is the 
chemical potential for electrons in band $\a$.  
Charge neutrality 
is guaranteed by $\m_{v}=-\m_{c}=-\d\m/2$ since 
$\hat{H}$ is particle-hole symmetric (the BI ground state is recovered 
for $\d\m=0$).
Exploiting the 
translational invariance, the mean-field equations for 
$\hat{H}_{\rm NEQ-GC}$ can be written as~\cite{footnote_TNEQEI}
\be
\left(
\begin{array}{cc}
    -\e_{k} +\d\m/2 & \D_{k} \\ \D_{k} & \e_{k}-\d\m/2
\end{array}
\right)  \left(
\begin{array}{c}
    \varphi^{\c}_{vk}  \\ \varphi^{\c}_{ck} 
\end{array}
\right) =  \c e_{k} \left(
\begin{array}{c}
     \varphi^{\c}_{vk}  \\ \varphi^{\c}_{ck} \end{array}
\right) ,
\label{hk(z)}
\ee
where $\e_{k}=[2V(1-\cos(k))+\e_{g}/2]$, $k\in(-\p,\p)$ is the 
quasimomentum, $\c=\pm$ 
labels the two eigensolutions and 
$\D_{k}=-\sum_{q}\tilde{U}_{k-q} b_{q}$ is the excitonic order parameter, $\tilde{U}_{k}$ 
being the Fourier transform of $U_{ij}$ and $b_{q}\equiv \varphi^{-}_{cq}  
\varphi^{-\ast}_{vq}$ the EC density. Only states of the minus 
branch are occupied since $e_{k}= \sqrt{(\ve_{k}-\d 
\m/2)^{2}+\D_{k}^{2}}\geq 0$. Equation~(\ref{hk(z)}) has to be solved 
self-consistently and $\D_{k}\neq 0$ 
implies a symmetry-broken NEQ-EI state.
The system remains a BI ($\D_{k}=0$) for $\d\m<\e_{\mathrm{x}}^{s}$, 
as it should be~\cite{Yamaguchi_NJP2012,PSMS.2019}.
A unique solution $\D_{k}=\D_{-k}$ (even in $k$) exists for 
$\e_{\mathrm{x}}^{s}<\d \m <\e_{\mathrm{x}}^{p}$ ($s$-wave NEQ-EI). 
For $\d\m>\e_{\mathrm{x}}^{p}$ we can find a solution $\D_{k}=\D^{s}_{k}+ 
\D^{p}_{k}$ with $\D^{s}_{k}$ ($\D^{p}_{k}$) 
even (odd) in $k$ for any fixed angle $\th={\rm 
arctan}(\D^{s}_{\p}/\D^{p}_{\p})\in(0,2\p)$.
The $p$-wave NEQ-EI state is realized when $\th=0$  and hence 
$\D_{k}=-\D_{-k}$. Below we show that this state can be generated by 
suitable laser pulses provided that 
the $p$-wave ($s$-wave) exciton is bright (dark).

\begin{figure}[tbp]
\includegraphics[width=7.35cm]{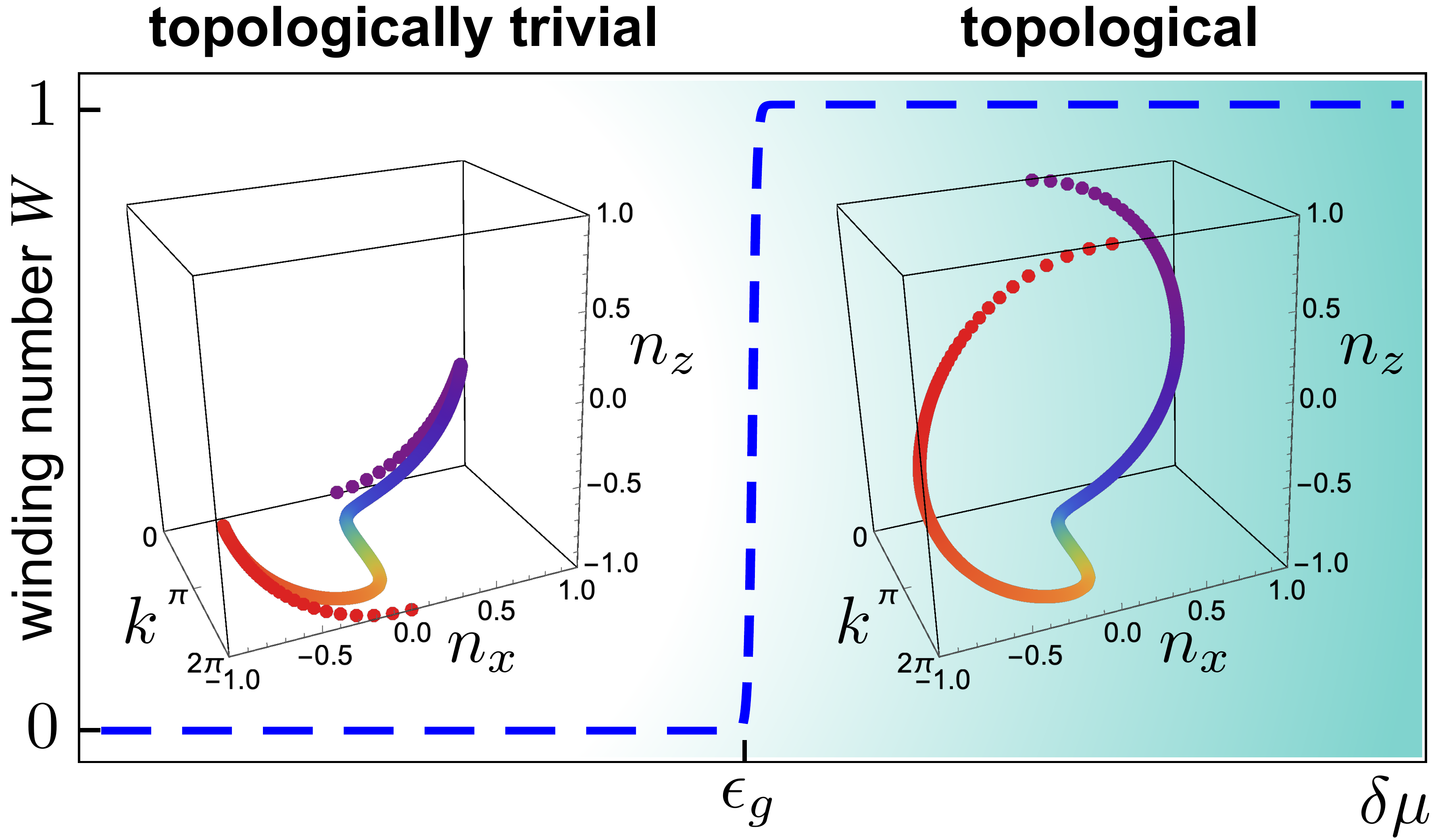}
\caption{
Winding number $W$ as a function of $\d \m$ (dashed line).
In the inset we show the path of the vector $\vec{d}_{k}$ 
in the nontopological phase for  $\d \m =0.96 <\e_{g}$ (left panel)
and in the topological phase for  $\d \m =1.04  >\e_{g}$.
All energies are in units of 
$\e_{g}$ and the function $\D_{k}$ is determined self-consistently 
for each $\d\m$.
}
\label{fig1}
\end{figure}

Independently of the symmetry the NEQ-EI state evolves according to the 
time-dependent mean-field equations $i\frac{d}{dt}\vf^{\c}_{k}(t)=
h^{\rm MF}_{k}(t)\vf^{\c}_{k}(t)$ where 
\be
h^{\mathrm{MF}}_{k}(t)=\left(
\begin{array}{cc}
    -\e_{k}  & \D_{k}(t) \\ \D^{\ast}_{k}(t) & \e_{k}
\end{array}
\right) 
\label{hhf(t)}
\ee
is the physical mean-field Hamiltonian. The excitonic order parameter
$\D_{k}(t)=\sum_{q}\tilde{U}_{k-q} \varphi^{-}_{cq} (t) 
\varphi^{-\ast}_{vq}(t)$ acquires a dependence on time through the 
wavefunctions. In Ref.~\cite{PSMS.2019} we have 
shown that this dependence is {\em monochromatic} and given by
\be
\D_{k}(t)=\D_{k}e^{i\d \m t}.
\label{delta(t)}
\ee
Thus, the mean-field hamiltonian supports self-sustained Josephson-like 
oscillations driven by the broken gauge symmetry. We then construct the
Floquet hamiltonian $h^{\mathrm{Floq}}_{k}$ from $\mathcal{T}e^{ -i \int_{0}^{T} 
dt\,h^{\mathrm{MF}}_{k}(t)} = e^{-ih^{\mathrm{Floq}}_{k} T}$,
where $\callT$ is the time-ordered operator and $T=2\pi/\d\m$, and 
look for nonvanishing Floquet topological invariants.
Since $h^{\mathrm{MF}}_{k}(t)$ is a $2\times 2$ monochromatic and 
hermitian matrix the Floquet hamiltonian can easily be 
calculated~\cite{PS.2015}
\be
h^{\mathrm{Floq}}_{k}=\left(
\begin{array}{cc}
    -\e_{k}  & \D_{k} \\ \D_{k}  & \e_{k} -\d \m
\end{array}
\right) = -\frac{\d \m}{2} \mathds{1} +  e_{k} \, \vec{n}(k)\cdot 
\vec{\s}
\label{heff}
\ee
where $\s_{x,y,z}$ are the Pauli matrices
and $\vec{n}(k)=\{n_{x}(k),n_{y}(k),n_{z}(k)\}=\{\D_{k}/e_{k},0,(\d 
\m/2-\e_{k})/e_{k}\}$. Interestingly, $h^{\mathrm{Floq}}_{k}$ 
coincides with the NEQ gran-canonical mean-field hamiltonian in 
Eq.~(\ref{hk(z)}) up to a constant diagonal shift.
The  winding number~\cite{Niu_PhysRevB.85.035110,Tong_PhysRevB.87.201109,Thakurathi_PhysRevB.88.155133} 
\be
W=\frac{1}{2\pi}\int_{-\pi}^{\pi}d\Theta_{k} \quad , \quad  
\Theta_{k}=\arctan \frac{n_{z}(k)}{n_{x}(k)}.
\label{winding}
\ee
measures the number of windings of the unit vector $\vec{d_{k}}=\{n_{x}(k),n_{z}(k)\}$
as $k$ crosses the first Brillouin zone. 
$W$ is a positive or negative integer in the 
topological phase and it is otherwise zero.
It is immediate to realize that $W=0$ 
for any even $\D_{k}$. 
If, instead,  $\D_{k}$ is an odd function then $W=\pm 1$ provided that 
$\d \m > \e_{g}$, see 
Fig.~\ref{fig1}. Thus a topological transition occurs
in a {\it p-}wave NEQ-EI as $\d\m$, and hence the average EC density 
$b=\frac{1}{\mathcal{N}}\sum_{q}b_{q}$ (with $\callN$  the number of cells),
exceeds a critical value. 
In Fig.~\ref{fig1} we also show the path of $\vec{d}_{k}$ resulting 
from the self-consistent solution of Eq.~(\ref{hk(z)}).
The difference in chemical potentials is $\d \m =0.96$ (left 
panel) and $\d \m =1.04$ (right panel).

\begin{figure}[tbp]
\includegraphics[width=7.35cm]{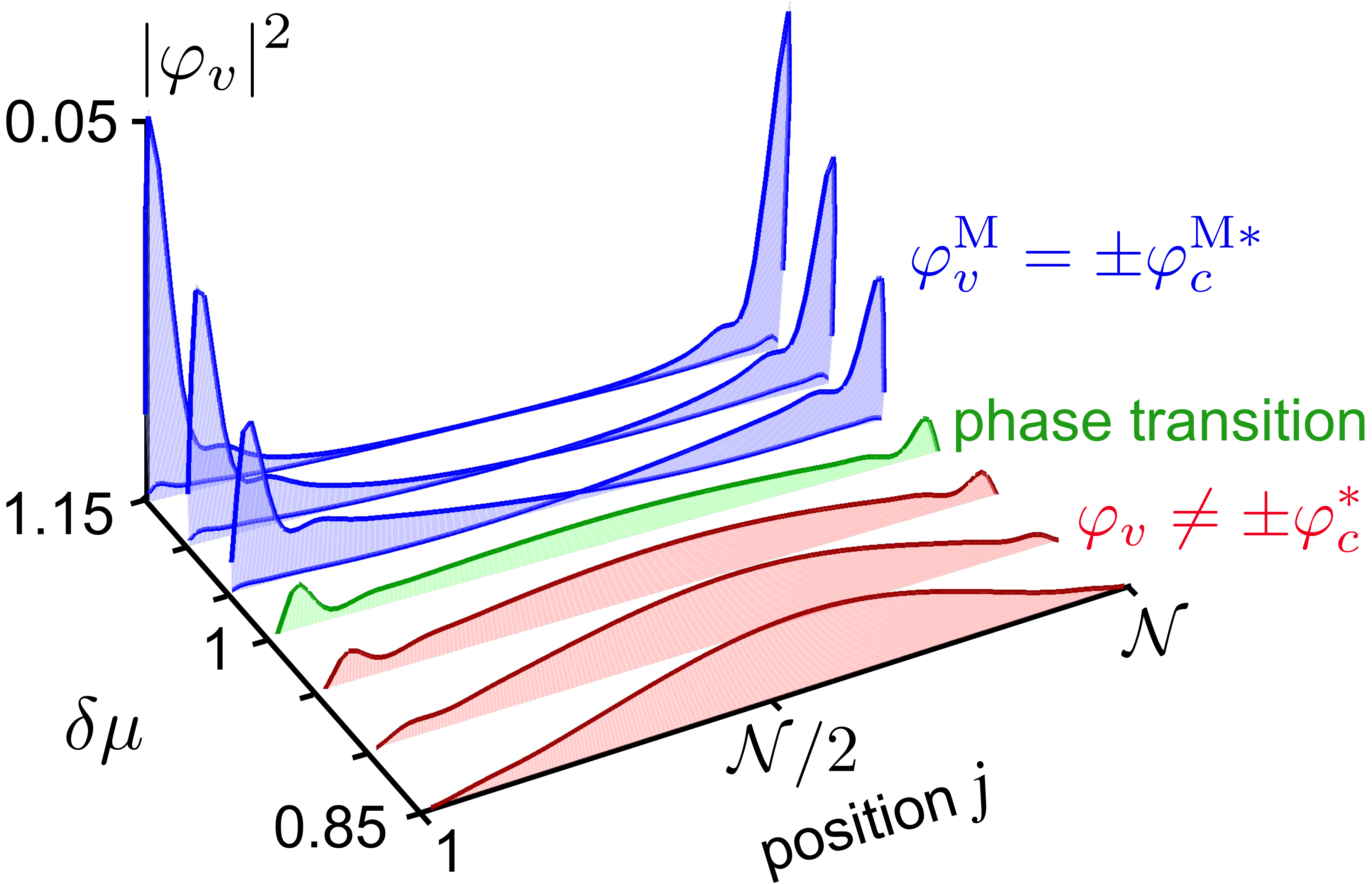}
\caption{Square modulus of the valence component of the 
eigenfunctions of the mean-field NEQ gran-canonical Hamiltonian with 
eigenvalue $e_{\rm max}$. For $\d\m<\e_{g}$ the eigenvalue $e_{\rm 
max}$ is strictly negative  and nondegenerate. For $\d\m>\e_{g}$ the eigenvalue $e_{\rm 
max}=0$ and the degeneracy is two. The corresponding eigenfunctions 
are Majorana modes. Energies are 
in units of $\e_{g}$.}
\label{fig2}
\end{figure}
 
According to the bulk-edge correspondence, a number $|W|$ of topologically 
protected Floquet Majorana modes should form at each open 
boundary~\cite{Tong_PhysRevB.87.201109}. As the the Floquet 
hamiltonian in Eq.~(\ref{heff}) coincides with the mean-field NEQ gran-canonical 
hamiltonian in Eq.~(\ref{hk(z)})  we 
consider $\hat{H}_{\rm NEQ-GC}$ on an open wire of $\callN=100$ 
cells and solve the mean-field equations in the site basis.
The spectrum is symmetric around zero energy with positive 
and negative eigenvalues $e^{+}_{\l}=-e^{-}_{\l}\geq 0$.  For $\d 
\m<\e_{g}$ the maximum energy $e_{\rm max}={\rm 
max}_{\l}\{e^{-}_{\l}\}$ is strictly negative and the eigenfunctions 
$\vf^{\pm }_{\rm max}$ of energies $\pm e_{\rm max}$ are delocalized 
along the wire. In Fig.~\ref{fig2} we plot the valence probability 
$|\vf^{+}_{{\rm max},vj}|^{2}$ versus site $j$ (the conduction 
probability $|\vf^{-}_{{\rm max},cj}|^{2}$ is identical). A 
sharp transition occurs for $\d \m>\e_{g}$ since the spectrum is 
almost the same as for  $\d \m<\e_{g}$ except for two degenerate eigenvalues 
appearing at zero energy. The corresponding eigenfunctions can be 
chosen to satisfy $\vf^{\rm M}_{vj}=\pm\vf_{cj}^{\rm M\ast}$, i.e., they are Majorana 
modes, and their valence components are plot in Fig.~\ref{fig2}. The Majorana 
modes are correctly localized at the system boundaries and the degree 
of localization 
increases as $\d \m$ moves deeper inside the topological phase.

\begin{figure}[tbp]
\includegraphics[width=7.35cm]{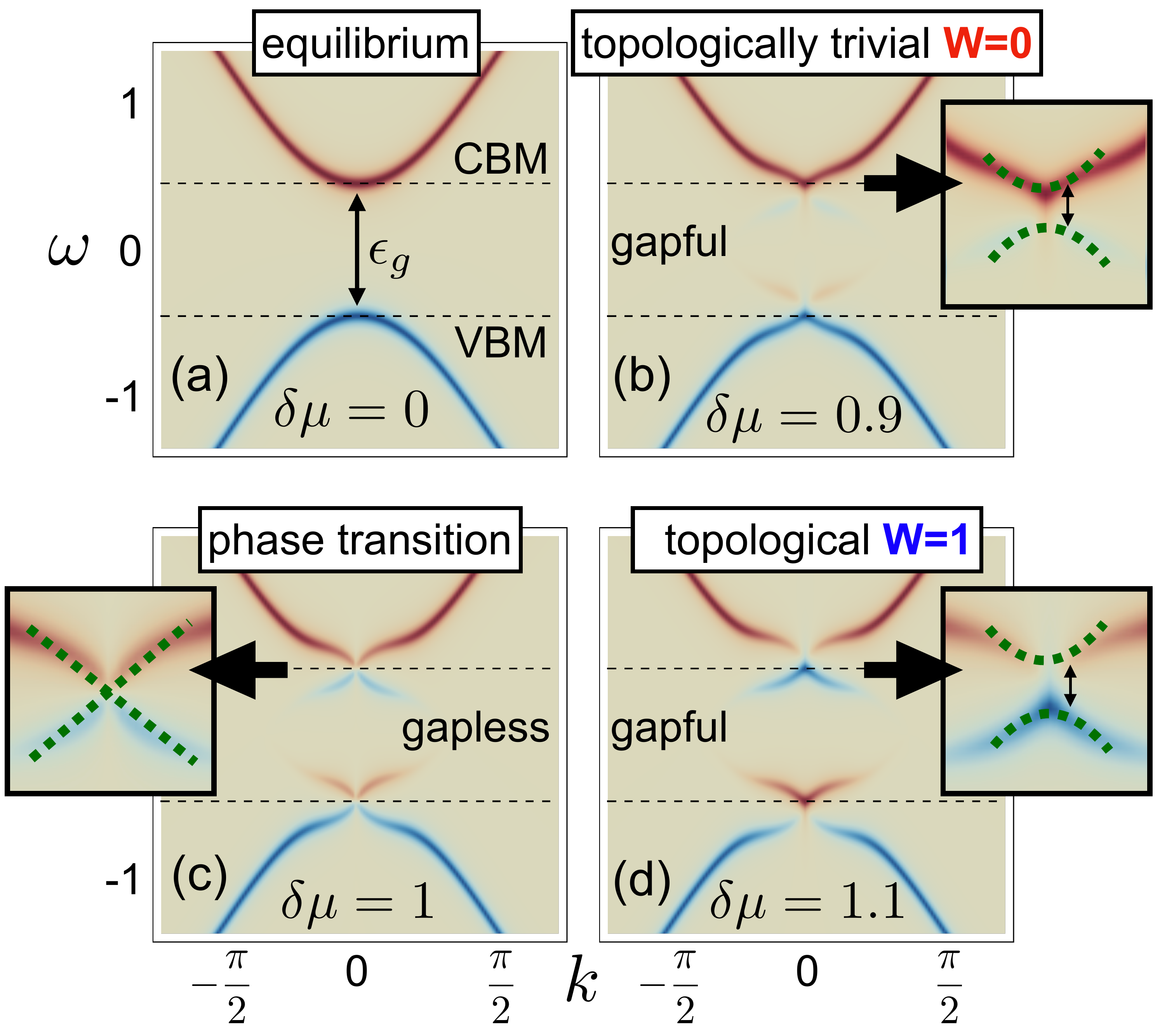}
\caption{
Contour plot of the spectral function $A_{k}(\w)$ 
for different values of $\d \m$. The red (blue) color refers to the 
removal contribution $A^{<}_{k}(\w)$ (addition contribution 
$A^{>}_{k}(\w)$).
The delta functions in Eq.~(\ref{spectral}) have been 
approximated by lorentzians of width $\eta=0.05$. The insets 
show a magnification of the spectral region around the conduction 
band maximum at $\e_{g}/2$. All 
energies are in units of $\e_{g}$.
}
\label{fig3}
\end{figure}
\begin{figure}[tbp]
\includegraphics[width=7.35cm]{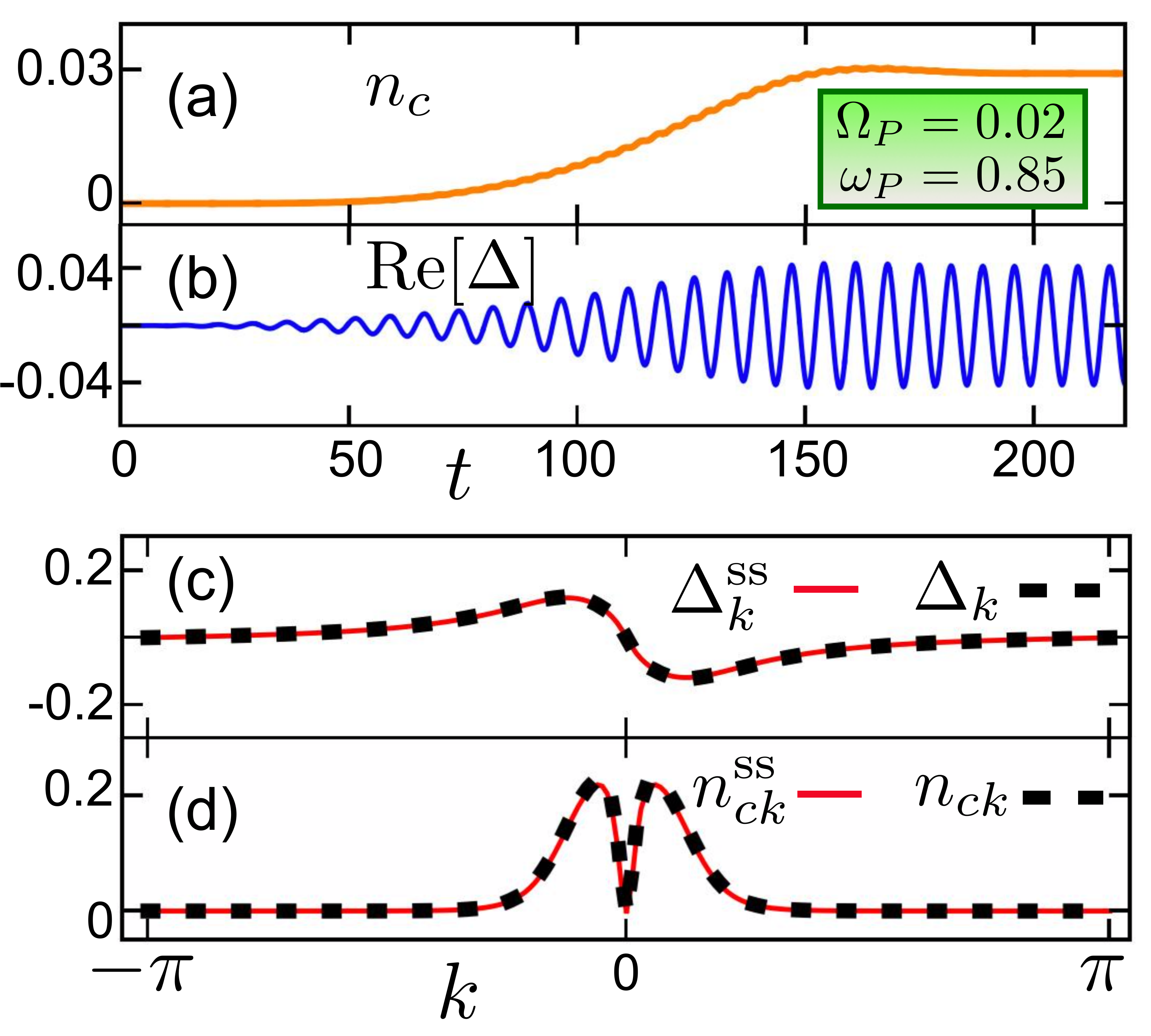}
\caption{
Time evolution of the conduction density $n_{c}$ (panel~a) and 
the real part of the averaged order parameter $\D$ (panel~b) induced 
by the laser pulse in Eq.~(\ref{pulse})
with $\W_{P}=0.02$, $\w_{P}=0.85$ and $T_{P}=200$.
Comparison of $\D^{\mathrm{ss}}_{k}$ and $n^{\mathrm{ss}}_{ck}$ 
(panels~c-d) extracted from the real-time simulation (solid red lines) with the 
self-consistent $\D_{k}$ and  $n_{ck}$ 
obtained from Eq.~(\ref{hk(z)}) with $\d \m =\d \m^{\rm ss}=0.907$    (dashed black lines).
Energies are in units of $\e_{g}$, and times in units of 
$1/\e_{g}$.
}
\label{fig4}
\end{figure}
\begin{figure}[tbp]
\includegraphics[width=7.35cm]{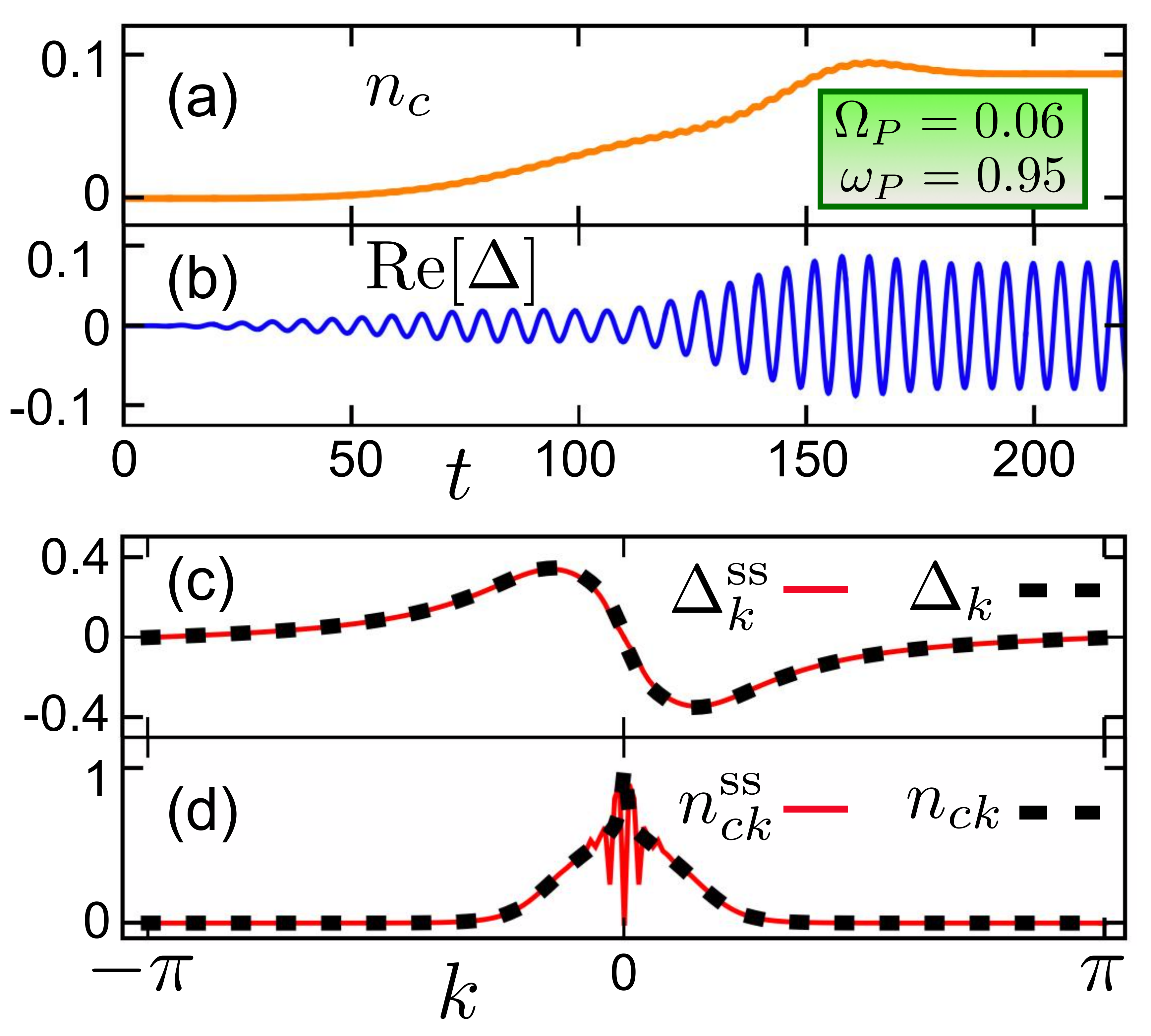}
\caption{Same as in Fig.~\ref{fig4} but with  laser
parameters  $\W_{P}=0.06$, $\w_{P}=0.95$ and $T_{P}=200$. For the 
self-consistent solution of Eq.~(\ref{hk(z)}) we have used $\d \m =1.04$.  }
\label{fig5}
\end{figure}

The topological  transition in a (bulk) {\it p-}wave NEQ-EI leave unique fingerprints 
on the ARPES spectrum
The spectral function $A_{k}(\w)$ is the sum of a 
removal ($<$)  and addition ($>$) contribution, 
$A_{k}(\w)=A^{<}_{k}(\w)+A^{>}_{k}(\w)$ with
\begin{align}
A^{\gtrless}_{k}(\w)&=|\varphi^{\pm}_{v 
k}|^{2}\d(\w \mp e_{k} +\frac{\d \m}{2})+|\varphi^{\pm}_{c 
k}|^{2}\d(\w \mp e_{k} - \frac{\d \m}{2}).
\label{spectral}
\end{align}
In Fig.~\ref{fig3} we show how $A_{k}(\w)$ changes
from the equilibrium (panel~a) to the 
 symmetry-broken (panel~b) and 
topological phase (panels~c-d).
As $\d \m$ overcomes  $\e_{\mathrm{x}}^{p} =0.82$ the system becomes 
a nontopological NEQ-EI 
and an excitonic structure appears inside the gap 
(panel~b)~\cite{PSMS.2016,PSMS.2019}.
For $\d \m =0.9<\e_{g}$ the conduction density is 
$n_{c}=\frac{1}{\mathcal{N}}\sum_{k}|\varphi^{-}_{ck}|^{2}=0.03$ and the  
averaged order parameter $\D\equiv \frac{2}{\mathcal{N}}\sum_{0<k<\p}\D_{k}=0.04$ 
(with $\callN$ the number of cells).
The removal (blue) component of the excitonic structure is separated 
from the bottom of the conduction band (red) by a small gap, 
consistently with the insulating
character of the state (see inset of Fig.~\ref{fig3}~b).
In contrast with the {\it s-}wave NEQ-EI~\cite{PSMS.2019,PBS.2020},
however, the excitonic structure has a vanishing spectral weight 
around the $\G$-point since $n_{c k}=|\varphi^{-}_{ck}|^{2}$ vanishes at 
$k=0$ in the nontopological  
phase (see also the dashed line in Fig.~\ref{fig4}~d).
At the topological critical point ($\d \m = \e_{g}$) we find $n_{c}=0.07$ 
and $\D=0.06$ (both larger than for $\d \m =0.9$). Despite $\D\neq 0$ the gap closes 
(see inset of Fig.~\ref{fig3}~c)
and the dispersion of the excitonic structure around the $\G$-point 
becomes, see Eq.~(\ref{spectral}), 
$E_{k}=\frac{\d \m}{2}- e_{k}\approx \frac{\d \m}{2}- \g |k|$
where we have approximated $\D_{k}\approx \g k$ 
(see also the dashed line in Fig.~\ref{fig4}~c and Fig.~\ref{fig5}~c).
Thus, the system becomes a Dirac semimetal. The transition point is 
also characterized by a discontinuity in $n_{c k=0}$ which 
varies abruptly from 0 ($\d\m<\e_{g}$) to 1 ($\d \m>\e_{g}$).
In the topological phase $\d \m=1.1>\e_{g}$ both 
$n_{c}=0.11$ and $\D=0.08$ increase and the gap re-opens,  
see Fig.~\ref{fig3}~d. Noteworthy, the spectral weight of the excitonic structurelorentzians
is now largest at the $\G$-point due to the aforementioned 
discontinuity in $n_{c k=0}$.

The remaining issue to be addressed is whether and how
the topological {\it p-}wave NEQ-EI state can be prepared.
We answer affirmatively provided that the {\it p-}exciton is 
much brighter than the {\it s-}exciton.
We consider the system initially in the BI ground state and 
drive it out of equilibrium by a laser pulse
\be
\hat{H}_{\rm laser}(t)=E(t)\sum_{k}D_{k}(\hat{\q}^{\dag}_{ck}\hat{\q}_{vk}+\hat{\q}^{\dag}_{vk}\hat{\q}_{ck}),
\label{Hdrive}
\ee
where $D_{k}$ is the valence-conduction dipole moment.
The electric field $E(t)$ is a pulse of finite duration $T_{P}$
centered around frequency $\w_{P}$:
\be
E(t)=\th(1-|1-2t/T_{P}|)
E_{P}\sin^{2}(\frac{\p t}{T_{P}})\sin(\w_{P}t).
\label{pulse}
\ee
To enhance absorption by the $p$-exciton we take $D_{k}$ odd in $k$. 
First we generate the {\it p-}wave NEQ-EI in the nontopological 
phase by tuning 
the  Rabi frequency $\W_{P}\equiv 
E_{P}D$  and the central frequency $\w_{P}$ in the range 
$(\e_{\mathrm{x}}^{p},\e_{g})$. 
In Fig.~\ref{fig4} we show the outcome of a real-time mean-field simulation 
performed with the CHEERS code~\cite{PS-cheers} using $D_{k}=D\sin (k)$ 
and optimal laser parameters 
$T_{P}=200$, $\W_{P}=0.02$ and $\w_{P}=0.85$
(times in units of $1/\e_{g}$).
At the end of the pulse
the system has steady-state occupations $n^{\rm ss}_{ck}$ giving 
a density $n_{c}= 0.03$ (panel a) and an averaged 
order parameter oscillating in time
as $\D(t)=\D^{\rm ss} e^{i\d \m^{\rm ss}t}$ with steady-state 
amplitude $\D^{\rm ss}=0.02$ and $\d 
\m^{\rm ss}=0.907$  (panel b). This oscillatory behavior 
is actually found for each $k$,  i.e. 
$\D_{k}(t)=\D^{\mathrm{ss}}_{k}e^{i\d \m^{\rm ss}t}$
(not shown). We mention that the self-sustained 
oscillations (persistent at the mean-field level) 
survive for long time when numerically exact propagation schemes are 
used~\cite{murakami2019ultrafast}.
In Fig.~\ref{fig4}~c-d we compare  $\D_{k}$ and $n_{ck}$
obtained from the self-consistent solution 
of  Eq.~(\ref{hk(z)}) at $\d \m=\d 
\m^{\rm ss}<\e_{g}$ against 
 $\D^{\rm ss}_{k}$ and $n^{\rm ss}_{ck}$ obtained from the real-time 
simulation. The agreement is remarkably good, thereby proving that 
a dynamical phase transition from a BI to a 
nontopological {\it p-}wave NEQ-EI can be induced by properly choosing 
the laser pulse.

Less obvious is the possibility of driving the BI toward a topological {\it 
p-}wave NEQ-EI. The BI ground state has $W=0$ and hence 
the system should experience a {\it dynamical} topological transition 
featuring a gap closure at the quantum critical point.
According to the Kibble-Zurek mechanism~\cite{Kibble_1976,Zurek1985} 
this introduces a diverging time-scale preventing the realization of 
the topological target state with an external field of 
{\it finite duration} $T_{P}$~\cite{Perfetto_PhysRevLett.110.087001,Sacramento_PhysRevE.90.032138}. 
In fact, topological defects are produced around the $\G$-point 
(gap-closure point) for any $T_{P}<\iif$~\cite{Sengupta_PhysRevLett.100.077204,Mondal_PhysRevB.78.045101,Bermudez_PhysRevLett.102.135702,Bermudez_2010,Lee_PhysRevB.92.035117,Defenu_PhysRevB.100.184306}.
One can easily show that for odd dipoles $D_{k}$ 
the density $n_{ck=0}(t)=0$ at every time 
(in the topological phase $n_{ck=0}=1$).
In Fig.~\ref{fig5} we show the results of a mean-field real-time 
simulation for a laser pulse with $T_{P}=200$,
$\W_{P}=0.06$ and $\w_{P}=0.95$.
As expected, the conduction density (panel a) and the amplitude 
of the averaged order 
parameter (panel b) are larger than in the previous case, see 
Fig.~\ref{fig4}. Of more importance is that they both attain a 
steady-state value after the pulse ($t>T_{P}$)  and that
(with high numerical accuracy)
the entire order parameter oscillates monochromatically 
as $\D_{k}(t)=\D^{\mathrm{ss}}_{k} e^{i\d \m^{\rm ss}t}$ with 
$\d\m^{\rm ss}=1.04>\e_{g}$. As Fig.~\ref{fig5}~c shows, 
$\D^{\mathrm{ss}}_{k}$ is indistinguishable from the self-consistent $\D_{k}$ of 
Eq.~(\ref{hk(z)}) at $\d\m=\d\m^{\rm ss}$. We conclude that for 
$t>T_{P}$ the mean-field hamiltonian has the same form as 
Eqs.~(\ref{hhf(t)},\ref{delta(t)}); hence the corresponding 
Floquet hamiltonian has winding number $W=1$. Although the two 
hamiltonians are identical the time-dependent state differs from the 
self-consistent one. This is shown in Fig.~\ref{fig5}~d where 
a difference  between $n_{ck}^{\rm ss}$ and the self-consistent 
$n_{ck}$ is clearly visible around $k=0$.
Such topological defect, however, does 
not destroy the topological phase since the Floquet hamiltonian 
depends only on $\D_{k}$.

To summarize, we have shown the existence of a Floquet topological 
phase in {\em nondriven} nonequilibrium matter  and how to steer a 
nontopological BI toward this phase with laser pulses of finite duration.
The nontrivial topology  emerges exclusively from the electron-hole 
Coulomb attraction and it leaves unique fingerprints in the ARPES 
spectra.
Our discussion is based on a paradigmatic 1D model, but the results 
are general and can easily be extended to 2D systems. In this case 
the $p_{x}+ip_{y}$ symmetry of the EC order parameter can be 
exploited to generate a  
nonvanishing Chern number $\frac{1}{4\pi}\int d^{2}k\, \vec{n}\cdot 
(\partial_{k_{x}}\vec{n}\times \partial_{k_{y}}\vec{n} )$ that, 
again, counts the Majorana edge modes.
Materials with optically bright $p$-excitons 
for realizing the topological {\it p-}wave NEQ-EI phase 
may include
semiconducting nanotubes~\cite{Maultzsch_PhysRevB.72.241402,Verdenhalven_2013}, biased 
graphene bilayers~\cite{ParkNL10_2010}, and low-dimensional compounds with strong spin-orbit coupling~\cite{Garate_PhysRevB.84.045403}.
An {\it s-}wave NEQ-EI phase has been recently 
observed in  bulk GaAs~\cite{Murotani_PhysRevLett.123.197401}: 
the way to light-induced topological phases of NEQ-EI has 
therefore been opened.

 \vspace{1cm}
 
{\it Akcknowledgements} We acknowledge useful discussions with
Andrea Marini and Davide Sangalli. We also acknowledge funding from
MIUR PRIN Grant No.20173B72NB and from INFN17-Nemesys project.


\begin{thebibliography}{52}
\bibitem[{foo()}]{footnote_TNEQEI}
\bibinfo{note}{The mean-field Hamiltonian in Eq.~(\ref{hk(z)}) in principle
  contains a renormalization of the bare bands $\e_{\a k}$ due to the Hartree
  self-energy. In this work we omit for simplicity such correction that,
  however, can be readsorbed in the defintion of the chemical potentials
  $\m_{\a}$.}
\end{thebibliography}

\end{document}